\documentclass[smallextended]{svjour3}
\usepackage{graphicx}
\usepackage{amsmath,amssymb}
\newcommand{\p}[2]{\frac{\partial #1}{\partial #2}}
\newcommand{\pp}[3]{\frac{\partial^2 #1}{\partial #2\partial #3}}
\newcommand{\pnq}[3]{\frac{\partial^{#3} #1}{\partial #2^{#3}}}
\newcommand{\dnq}[3]{\frac{d^{#3} #1}{d #2^{#3}}}
\newcommand{\de}[2]{\frac{d #1}{d #2}}

\def\calq{{\cal Q}}

\title{Bounds on Surface Stress Driven Shear Flow}

\author{George I. Hagstrom \and Charles R. Doering}

\institute{George I. Hagstrom \at Magneto-Fluids Division, Courant Institute of Mathematical Sciences, New York University, New York, NY, 10012-1185 USA \\\email{hagstrom@cims.nyu.edu}
\and Charles R. Doering \at Departments of Physics and Mathematics, and Center of the Study of Complex Systems, University of Michigan, Ann Arbor, Michigan, 48109-1034 USA \\\email{doering@math.umich.edu}}
\date{}
\begin{document}

\titlerunning{Stress Driven Shear Flow}

\authorrunning{G. I. Hagstrom and C. R. Doering}

\maketitle


\bigskip



\begin{abstract}
The background method is adapted to derive rigorous limits on surface speeds and bulk energy dissipation for shear stress driven flow in two and three dimensional channels.
By-products of the analysis are nonlinear energy stability results for plane Couette flow with a shear stress boundary condition: when the applied stress is gauged by a dimensionless Grashoff number $Gr$, the critical $Gr$ for energy stability is $139.5$ in two dimensions, and $51.73$ in three dimensions.
We derive upper bounds on the friction (a.k.a. dissipation) coefficient $C_f = \tau/\overline{u}^2$, where $\tau$ is the applied shear stress and $\overline{u}$ is the mean velocity of the fluid at the surface, for flows at higher $Gr$ including developed turbulence: $C_f \le 1/32$ in two dimensions and $C_f \le 1/8$ in three dimensions.  This analysis rigorously justifies previously computed numerical estimates.
\end{abstract}

\section{Introduction}

One of the great challenges facing modern mathematical physics and applied mathematics is to deduce turbulent transport properties directly from the fundamental equations of motion, often the Navier-Stokes equations describing the flow of incompressible Newtonian fluids.
This problem remains unsolved in general, but recent decades have witnessed significant progress in the derivation of rigorous estimates of complex flow characteristics.

One approach to the analysis is the so-called ``background method'' based on a decomposition of the velocity (or temperature) field into a steady incompressible component that absorbs the inhomogeneous boundary conditions maintaining the flow and the associated dynamical fluctuations \cite{CDPROLA}.
Roughly speaking, when the background component satisfies what appears to be a nonlinear energy stability condition as if it was a steady solution sustained by suitable forces (or heat sources), then it yields an upper bound on the actual transport of momentum (or heat) by all solutions with those boundary conditions.
The background method was first applied to bound turbulent dissipation in high Reynolds number shear flows driven by boundary motion, i.e., for the traditional plane Couette geometry and boundary conditions where the velocity field satisfies inhomogeneous Dirichlet boundary conditions \cite{CDPROLA,CD1994} .

In many applications, however, flows are driven by stresses, i.e., momentum (or heat) fluxes, on a surface.
Mathematically this means that driving Dirichlet conditions are replaced with inhomogeneous Neumann boundary conditions which presents some challenges for the background analysis: the fluctuations are no longer ``pinned'' to the boundaries and the stability-like character of backgrounds are correspondingly more difficult to establish.
This issue was first encountered in application of the background method to Rayleigh-B\'enard convection.
For conventional fixed-temperature conditions where temperature fluctuations vanish at the boundaries the background method produces rigorous upper limits to the heat flux \cite{CD1996,DOR2006,WD2011}.
If the heat flux at the boundaries is specified, however, temperature fluctuations at the boundaries are not so constrained.
Nevertheless the background analysis could be adapted to derive lower limits on temperature drop across the layer that correspond, in terms of the high-Rayleigh number scaling, with the fixed temperature bounds \cite{FixedFlux2002,Ralf}.

In physical oceanography applications it is natural to consider shear flows driven by (wind) stresses applied at a (top) surface, and this scenario presents a new set of challenges for the background method.
In this case the goal is to derive an estimate of the mean surface flow speed from which the statistically steady state bulk dissipation may be deduced.
Tang, Caulfield, and Young \cite{TCY} first used the background method to study this problem, but rather than imposing stress boundary conditions they considered a modified model wherein a body force is applied in a thin layer near the upper surface of the layer satisfying a homogeneous Neumann condition, by numerically solving the Euler-Lagrange equations for the optimal upper bounds producing relations between the ``applied stress'' and the mean surface speed and bulk dissipation.
(They also applied a clever analysis method to establish the scaling rigorously.) Hagstrom and Doering \cite{HagstromDoeringMarangoni} applied the background method to Marangoni convection, which is also a flow driven by a stress (proportional to the horizontal temperature gradient) at the upper boundary.

In this paper we establish bounds for the stress-driven problem by adapting the background method to this class of problems.
We derive scaling relations corresponding precisely to those of Tang {\em et al} with reasonable prefactors. This leads to an upper bound on the friction coefficient $C_f$ and consequently a lower bound on $\gamma=1/\sqrt{C_f}$, which is what appears in Tang {\em et al}.
In the next section we describe the setup and introduce the notion of energy stability in the two-space-dimensional setting.
The following section 3 presents energy stability analysis in three space dimensions, and the subsequent sections 4 and 5 contains application of the background method to stress-driven shear flow.

\section{Stress driven flow and energy stability in two dimensions}

\begin{figure}
\begin{center}
\includegraphics[scale=.4]{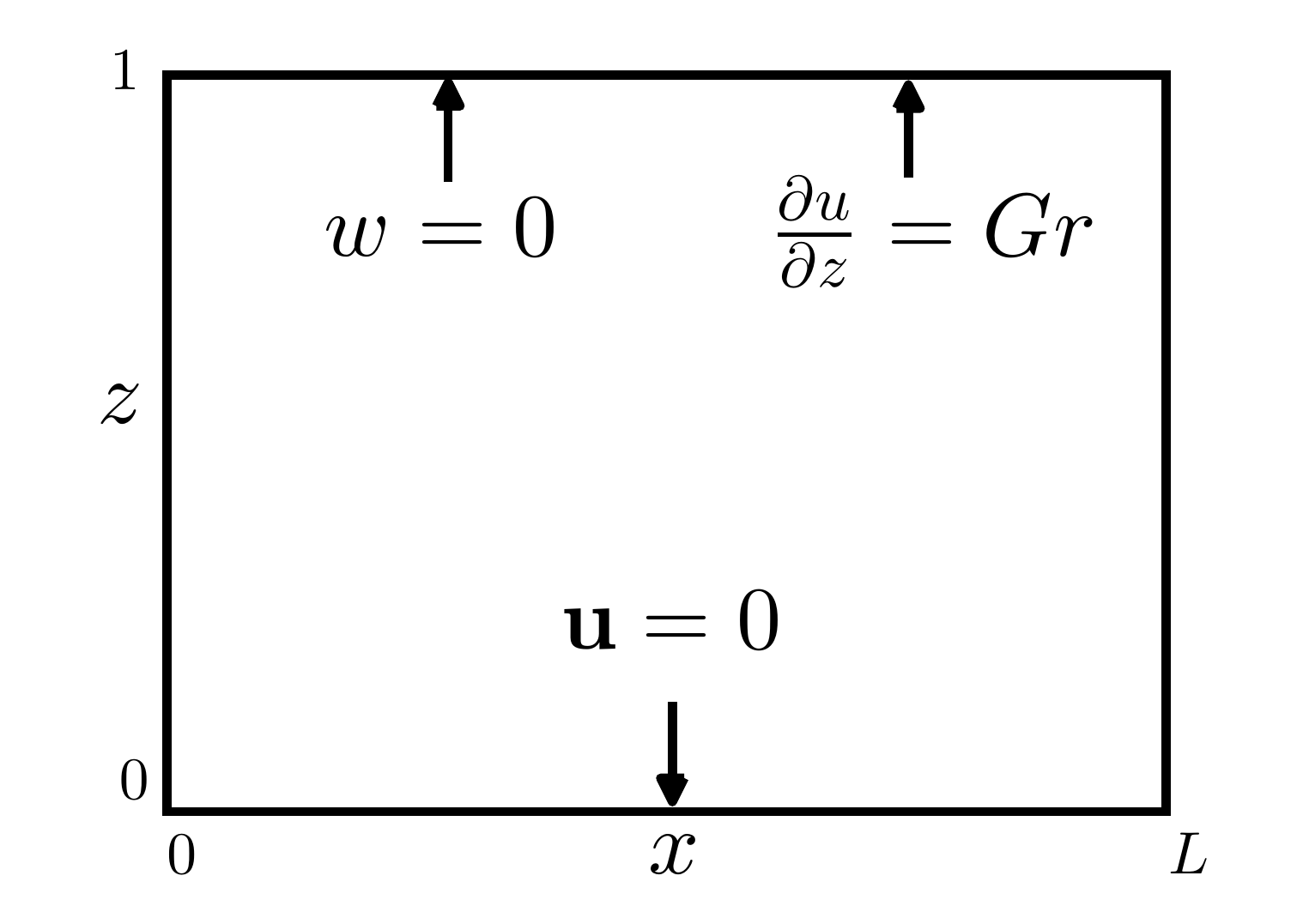}
\caption{Geometry for the 2d surface stress driven shear flow problem. Constant stress and no slip boundary conditions for ${\bf u}$ are shown at the upper and lower surfaces, and all dependent variables are periodic in $x$ with period $L$.}\label{fig:sheardomain1}
\end{center}
\end{figure}

Consider flow in the two dimensional domain shown in in Figure \ref{fig:sheardomain1} with periodic boundary conditions in the horizontal $x$ direction, a no slip condition on the bottom at $z=0$, and a fixed shear stress on top at $z=h$.
It is convenient to work with a nondimensional version of the system so we use the domain height $h$ as the length scale, $\frac{h^2}{\nu}$, where $\nu$ is the kinematic viscosity of the fluid, as the time scale, and $Gr=\frac{h^2\tau}{\nu^2}$ as the Grashoff number, where  $\tau$ is the stress applied at the upper surface of the fluid. The constant density throughout is scaled to unity via a suitable choice of mass units.
Then the equations of motion for the velocity vector field ${\bf u}({\bf x},t) = {\bf i}u(x,z,t) + {\bf k}w(x,z,t)$ and pressure $p(x,z,t)$ are 
\begin{align}
\frac{\partial {\bf u}}{\partial t}+{\bf u}\cdot\nabla {\bf u} +\nabla p &= \Delta {\bf u} \\
\nabla\cdot {\bf u} &= 0
\end{align}
with boundary conditions
\begin{equation}
{\bf u}|_{z=0} = 0, \quad w|_{z=1} = 0, \quad \frac{\partial u}{\partial z}|_{z=1} = Gr.
\end{equation}
The simplest steady laminar solution of these equations, which exists for all parameter values, is the uniform shear (Couette) flow with ${\bf u} = {\bf i} \, Gr \, z$ and $p = \ $ constant.
Energy stability theory ensures that this solution is nonlinearly stable at sufficiently low $Gr$.

The analysis begins by making the substitution ${\bf u} = {\bf i} \, Gr \, z + {\bf u'}$ denoting the fluctuations by ${\bf u'}={\bf i} \, u' + {\bf k} \, w'$.
The fluctuations' equations of motion are
\begin{align}
\p{{\bf u'}}{t} + {\bf u'}\cdot\nabla {\bf u'} + Gr \, z\p{{\bf u'}}{x} + {\bf i} \, Gr \, w' +\nabla p & = \Delta {\bf u'} \\
\nabla\cdot {\bf u'} &= 0
\end{align}
with homogeneous Dirichlet and Neumann boundary conditions
\begin{equation}
{\bf u}'|_{z=0} = 0, \quad \p{u'}{z}|_{z=1} = 0, \quad w'|_{z=1} = 0.
\end{equation}

The dot product of the momentum equation with ${\bf u'}$ and an integration over the domain, integrating by parts with the help of the homogeneous boundary conditions for the fluctuations, leads to the energy evolution equation
\begin{equation}
\frac{d}{dt} \frac{1}{2}\|{\bf u'}\|^2  = - \int  \left( |\nabla {\bf u'}|^2 + Gr \, u' \, w' \right) \, dx \, dz  \ \equiv \  -{\cal Q}\{ {\bf u'} \} 
\end{equation}
where $\| \cdot \|$ denotes the $L^2$ norm on the domain, and ${\cal Q}$ is a quadratic form defined by the above expression.
Energy stability theory is based on the observation that if ${\cal Q}\{ {\bf u'} \}$ is positive for all divergence-free ${\bf u'}$ satisfying the fluctuations boundary conditions, then Gronwall's lemma implies exponential decay of ${\|{\bf u'}(\cdot, t)\|^2}$ and thus unconditional stability of the base solution. 

The analysis proceeds by using variational methods to minimize $\calq$ subject to the constraints $\nabla\cdot {\bf u'}=0$ and $\|{\bf u'}\|^2=1$. The resulting Euler-Lagrange equations are the eigenvalue problem
\begin{align}
\lambda u' &= -\Delta u'+\frac{Gr}{2}w'+\p{q}{x} \\
\lambda w' &= -\Delta w'+\frac{Gr}{2}u'+\p{q}{z}
\end{align}
where the ``pressure'' $q({\bf x})$ is the Lagrange multiplier enforcing incompressibility.
If the smallest eigenvalue $\lambda_{min}>0$ then $\calq$ is positive definite and the base solution is stable.
To solve it we introduce the stream function $\Psi$ satisfing $\p{\Psi}{x}=w'$ and $\p{\Psi}{z}=-u'$ and eliminate the pressure to find the fourth order equation
\begin{equation} 
\lambda\Delta\Psi + \Delta^2\Psi + Gr \, \pp{\Psi}{x}{z} = 0
\end{equation}
with the four boundary conditions
\begin{equation}
\p{\Psi}{x}|_{z=0,1} = 0, \quad \p{\Psi}{z}|_{z=0} =0, \quad \pnq{\Psi}{z}{2}|_{z=1} = 0.
\end{equation}
The system is translation invariant in $x$ so we write $\Psi$ in terms of its Fourier series.
Writing $\Psi=\sum_k \hat{\Psi}_k e^{i \, k \, x}$, where $k = 2 \pi n/L $ for integer $n \in (\infty, \infty)$, the problem becomes the fourth-order ordinary differential equation
\begin{equation}
\lambda \left(\pnq{\hat{\Psi}}{z}{2}-k^2\hat{\Psi}\right) =
-\left(\pnq{\hat{\Psi}}{z}{4}-2k^2\pnq{\hat{\Psi}}{z}{2} + k^4\hat{\Psi}\right)- i \, Gr \,  k \, \p{\hat{\Psi}}{z}
\end{equation}
with
\begin{equation}
\hat{\Psi}|_{z=0,1} = 0, \quad \p{\hat{\Psi}}{z}|_{z=0} = 0, \quad \pnq{\hat{\Psi}}{z}{2}|_{z=1} = 0
\end{equation}
where, simply for notational neatness, we have suppressed the $k$ dependence of $\hat{\Psi}_k$.

We search numerically for the critical Grashoff number below which all of the eigenvalues $\lambda$ are positive. At the bifurcation point, where an eigenvalue first becomes
negative, $\lambda=0$ and $\hat{\Psi}$ must satisfy:
\begin{equation}
\left(\pnq{\hat{\Psi}}{z}{4}-2k^2\pnq{\hat{\Psi}}{z}{2} + k^4\hat{\Psi}\right)+ i \, Gr \,  k \, \p{\hat{\Psi}}{z}=0 \label{Gr2dGen}
\end{equation}

Therefore the critical Grashoff number is the smallest magnitude real generalized eigenvalue of (\ref{Gr2dGen}), where $Gr$ is the eigenvalue parameter. We discretized this eigenvalue problem using second order accurate finite differences with appropriate modifications to apply the boundary conditions, and used
Richardson extrapolation to accelerate the convergence of the resulting sequence of approximations to the smallest eigenvalue for
each value of $k$. The ultimate limitation on the accuracy of the computation was the condition number
of the differentiation matrix corresponding to $\pnq{\hat{\Psi}}{z}{4}$, which became extremely large as
the mesh was refined.
This approach leads us to conclude that the critical Grashoff number is at least $Gr = 139.54965$, and the value of $k$ where the first eigenvalue loses positivity is $k=3.146899$ (we note that $k$ seems very close to $\pi$). 

\section{Energy stability for three dimensional stress driven flow}

In three spatial dimensions the stress driven flow problem is
\begin{align}
\p{{\bf u}}{t}+{\bf u}\cdot\nabla {\bf u} +\nabla p&= \Delta {\bf u} \\
\nabla\cdot {\bf u} &= 0
\end{align}
with mixed Dirichlet and (inhomogeneous) Neumann conditions
\begin{equation}
w|_{z=0,1} = 0, \quad
u|_{z=0} = 0 = v|_{z=0}, \quad
\p{u}{z}|_{z=1} = Gr, \quad
\p{v}{z}|_{z=1} = 0.
\end{equation}
In three dimensions ${\bf u}({\bf x},t) = {\bf i} \, u(x,y,z,t) + {\bf j} \, v(x,y,z,t) + {\bf k} \, w(x,y,z,t)$ and the domain is periodic in both $x$ and $y$ with, respectively, periods $L_x$ and $L_y$.
Steady plane parallel Couette flow is again a solution and the same substitution ${\bf u}={\bf i} \, Gr \, z + {\bf u'} $ and analysis as in
the two dimensional case yields the Euler-Lagrange equations
\begin{align} 
\lambda u' &= -\Delta u'+\frac{Gr}{2}w'+\p{q}{x} \\
\lambda v' &= -\Delta v'+\p{q}{y} \\
\lambda w' &= -\Delta w'+\frac{Gr}{2}u'+\p{q}{z}\\
0 &= \nabla \cdot {\bf u'}
\end{align}

Assuming that the critical eigenfunction is independent of $x$, we can introduce a stream function $\Psi$ defined by $\p{\Psi}{y}=w'$ and $\p{\Psi}{z}=-v'$.
(The lowest eigenmodes of shear driven flows tend to be Langmuir-circulation-like flows, i.e., streamwise aligned rolls that are independent of the streamwise direction; this assumption is also justified by empirical observations \cite{TCY}.)
Then the Euler-Lagrange equations are 
\begin{align}
\lambda u'  &=\frac{Gr}{2}\p{\Psi}{y}-\Delta u' \\
-\lambda \p{\Psi}{z} &= \p{q}{y}+\Delta\p{\Psi}{z} \\
\lambda\p{\Psi}{y} &= \p{q}{z}-\Delta\p{\Psi}{y}+\frac{Gr}{2}u'
\end{align}
We then eliminate the pressure $q$ to obtain the system
\begin{align}
\lambda \p{u'}{y}  &=\frac{Gr}{2}\pnq{\Psi}{y}{2}-\Delta \p{u'}{y} \\
\lambda\Delta\Psi &=-\Delta^2\Psi+\frac{Gr}{2}\p{u'}{y}.
\end{align}
Finally we write ODEs for the Fourier modes of $u'$ and $\Psi$
\begin{align}
\lambda i \, k \, u' + \frac{Gr}{2} \, k^2 \, \hat{\Psi} + i \, k \, \left(\pnq{u'}{z}{2}-k^2u'\right) &= 0 \\
\lambda\left(\pnq{\hat{\Psi}}{z}{2}-k^2\hat{\Psi}\right) + \left(\pnq{\hat{\Psi}}{z}{4}-2k^2\pnq{\hat{\Psi}}{z}{2} +
k^4\hat{\Psi}\right) &=  \frac{Gr}{2} \, i \, k \, u' 
\end{align}
where $k = 2 \pi n/L_y $ for integer $n \in (\infty, \infty)$ with boundary conditions
\begin{equation}
\hat{\Psi}|_{z=0,1} = 0, \quad \p{\hat{\Psi}}{z}|_{z=0} = 0, \quad \pnq{\hat{\Psi}}{z}{2}|_{z=1} = 0, \quad u'|_{z=0} = 0, \quad \p{u'}{z}|_{z=1} = 0.
\end{equation}

This system can also be discretized and converted into a generalized eigenvalue problem exactly as before. The bifurcation occurs again through the loss of invertibility of the operator:
\begin{align}
 - i \, k \, \left(\pnq{u'}{z}{2}-k^2u'\right) &= \frac{Gr}{2} \, k^2 \, \hat{\Psi} \\
\left(\pnq{\hat{\Psi}}{z}{4}-2k^2\pnq{\hat{\Psi}}{z}{2} +
k^4\hat{\Psi}\right) &=  \frac{Gr}{2} \, i \, k \, u' \label{Gr3dGen}
\end{align}

We find the smallest magnitude generalized eigenvalue of (\ref{Gr3dGen}), using $Gr$ as the eigenvalue parameter. This was done by using second order centered differences combined with Richardson extrapolation. The critical Grashoff number is  $Gr=51.730001$, where the first negative eigenvalue appears at $k=2.085586$, in agreement with Tang {\emph et al} \cite{TCY} who imposed a body force in a vanishingly small layer near the upper surface to realize the shear stress boundary condition. Therefore the methods used in the respective 
papers suggest that flows driven by shear stress are similar to those driven by a body force in a narrow region near the upper surface in terms of their energy stability boundaries.

\section{Friction coefficient and bounds in two dimensions}

Define $\langle\cdot\rangle$ to be the space-time average and $\bar{\cdot}$ to be the horizontal-time average.
In dimensional variables the bulk energy dissipation rate, an emergent quantity (emergent meaning that it arises from complicated interactions of the individual constituents of the fluid) depending on the particular solution in this setup, is $\epsilon=\langle\nu|\nabla {\bf u}|^2\rangle$.
The Reynolds number, also an emergent quantity, is naturally defined $Re=\frac{\bar{u}(h)h}{\nu}$.
The familiar friction (dissipation) coefficient $C_f=\frac{\epsilon h}{\bar{u}(h)^3}$ is traditionally considered a function of $Re$; for the steady Couette solution $C_f(Re)=\frac{1}{Re}$.

For the system considered here the applied shear stress---$\tau$ in dimensional variables, $Gr$ nondimensionally---is the control parameter, so in order to express $C_f$ in the natural variables for analysis we need a connection between $\epsilon$, $\bar{u}(h)$, and $\tau$.
This comes from the global (mean) power balance: after taking the dot product of the momentum equation with ${\bf u}$ and averaging we find, in dimensional units, that
\begin{equation}
\epsilon = \left< \nu |\nabla {\bf u}|^2\right> = \frac{\tau \, \bar{u}(h)}{h}.
\end{equation}
Thus $C_f=\frac{\tau}{\bar{u}(h)^2} = \frac{Gr}{Re^2}$ and lower estimates for the (dimensional) mean surface speed $\bar{u}(h)$ as a function of $\tau$, i.e., lower bounds on $Re$ as a function of $Gr$, result in upper limits on the friction coefficient.
Properly adapted to the boundary conditions at hand, the background method may be employed to produce such lower bounds on Reynolds number as a function of the Grashof number.
We now turn to this analysis.

In the context of the non-dimensional equations, introduce a background horizontal velocity $U(z)$ satisfying $U|_{z=0} = 0$ and the inhomogeneous boundary condition $dU/dz|_{z=1}=Gr$. 
Write ${\bf u}={\bf i}U(z)+\tilde{{\bf u}}$ so that $\tilde{{\bf u}}$ solves
\begin{align}
\p{\tilde{{\bf u}}}{t}+\tilde{{\bf u}}\cdot\nabla\tilde{{\bf u}}+\nabla p + U \p{\tilde{{\bf u}}}{x}+{\bf i}\tilde{w} \de{U}{z} &= \Delta\tilde{{\bf u}}+{\bf i} \dnq{U}{z}{2} \label{eq:backgroundv}\\
\nabla\cdot\tilde{\bf{u}} &= 0
\end{align}
with homogeneous boundary conditions
\begin{equation}
\tilde{{\bf u}}|_{z=0} = 0, \quad \tilde{w}|_{z=1} = 0, \quad \p{\tilde{u}}{z}|_{z=1} = 0.
\end{equation}

Take the dot product with $\tilde{{\bf u}}$ and compute the space time average.
If the norm $\|{\tilde{\bf u}}\|$ is uniformly bounded in time, then $\left<d|\tilde{{\bf u}}|^2/dt\right> = 0$ and we see that:
\begin{align}
0&= -\left<|\nabla\tilde{{\bf u}}|^2\right>-\left<\de{U}{z}\tilde{u}\tilde
{w}\right>+\left<
\tilde{u}\dnq{U}{z}{2}\right> \\
&= -\left<|\nabla\tilde{{\bf u}}|^2\right>-\left<\de{U}{z}\tilde{u}\tilde{
w}\right>
+Gr(\bar{u}(1)-U(1))-\left<\p{\tilde{u}}{z}\de{U}{z}\right>.  \label{thing}
\end{align}


To establish uniform boundedness of $\|{\tilde{\bf u}}\|$ consider the space integral of the dot product of ${\tilde{\bf u}}$ with \ref{eq:backgroundv}:
\begin{equation}
\frac{1}{2}\de{}{t}\|{\tilde{\bf u}}\|^2 = -\|\nabla{\tilde{\bf u}}\|^2-\int dx dz \de{U}{z}{\tilde u}{\tilde w}+Gr\int dx {\tilde u}(1)
-\int dx dz \p{\tilde{u}}{z}\de{U}{z}
\end{equation}

Making use of the fundamental theorem of calculus for $\int dx {\tilde u}(1)$, and the Cauchy-Schwarz inequality we find:
\begin{align}
\frac{1}{2}\de{}{t}\|{\tilde{\bf u}}\|^2 &\leq  -\|\nabla{\tilde{\bf u}}\|^2-\int dx dz \de{U}{z}{\tilde u}{\tilde w} +\sqrt{L}Gr\|\nabla{\tilde u}\|
+\left\|\de{U}{z}\right\|\|\nabla{\tilde u}\| \\
&\leq -2 \calq_U\{\tilde{{\bf u}}\} +\sqrt{L}Gr\|\nabla{\tilde u}\|
+\left\|\de{U}{z}\right\|\|\nabla{\tilde u}\|
\end{align}

Here we have defined the quadratic form $\calq_U\{\tilde{{\bf u}}\}$ by:
\begin{equation}
\calq_U\{\tilde{{\bf u}}\} = \frac{1}{2}\|\nabla\tilde{{\bf u}}\|^2+\int dxdz \de{U}{z}\tilde{u}\tilde{w}
\end{equation}

\begin{figure}
\begin{center}
\includegraphics[width=2.5 in]{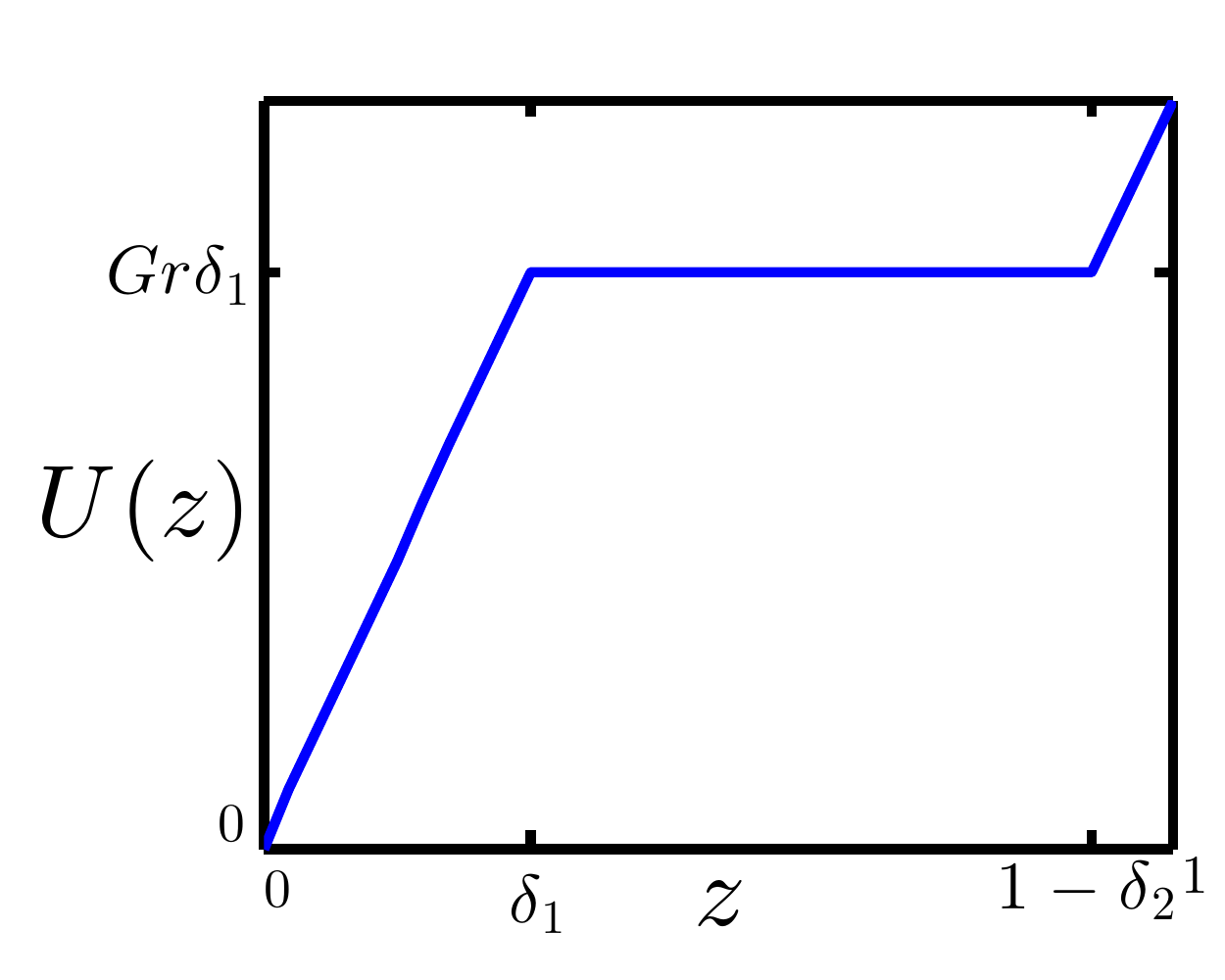}
\caption{Example of a background horizontal velocity profile $U(z)$, with boundary layers of width $\delta_1$ and $\delta_2$ where the slope satisfies $U'(z)=Gr$, and constant profile $U(z)=Gr\delta_1$ in the bulk region $\delta_1<z<1-\delta_2$.}\label{fig:background}
\end{center}
\end{figure}

We will choose $U$ so that we can bound $\calq_U\{\tilde{{\bf u}}\}$ by $C\|\nabla{\tilde u}\|^2$. We accomplish this by picking $U$ to
have a vanishing derivative in the bulk of the flow, only being non-zero in boundary layers so that the boundary conditions of the fluctuation field may be satisfied.
This may be accomplished by considering piece-wise linear background velocity profiles as shown in Figure \ref{fig:background} with boundary layers near the top and bottom of the layer.
That is, consider
\begin{equation}
U(z) = \left\{ \begin{array}{ccc}
			Gr \, z & \quad \textrm{for} \quad & 0<z<\delta_1     \\
			 Gr \, \delta_1    & \quad \textrm{for} \quad & \delta_1<z<1-\delta_2      \\
			Gr(\delta_1+\delta_2+z-1)  & \quad \textrm{for} \quad & 1-\delta_2<z<1 
					\end{array} \right. \label{background}.
\end{equation}

For notation convenience we henceforth drop the accents and refer to the fluctuations away from the background as ${\bf u} = {\bf i}u+{\bf k}w$.
Then
\begin{align}
\calq_U = \frac{1}{2}\|\nabla {\bf u}\|^2 &+ Gr \int_0^L dx \int_0^{\delta_1}  dz \ u(x,z) \, w(x,z) \nonumber \\
&+ Gr  \int_0^L dx \int_{1-\delta_2}^1  dz \ u(x,z) \, w(x,z). \label{Qd22}
\end{align}

To deduce acceptable values for $\delta_1$ and $\delta_2$, rewrite the second term on the right hand side of (\ref{Qd22}) above as
\begin{equation}
\int_0^L dx \int_0^{\delta_1}  dz \ u \, w =  \int_0^L dx \int_0^{\delta_1}  dz \int_0^z dz' \left(\p{u}{z'}w+u\p{w}{z'}\right).
\end{equation}
Incompressibility implies $u\p{w}{z'}=-u\p{u}{x}=-\frac{1}{2}\p{u^2}{x}$ which in turn implies that the second term in the integral above integrates to zero.

Thus, introducing the notation
\begin{equation}
\|h\|^2_{x(a,b)}=\int dx \int_a^b dz \,  h(x,z)^2, \label{note}
\end{equation}
using successive applications of the Cauchy-Schwarz and Young's inequalities, and recalling incompressibility once again, we deduce
\begin{align}
& \left| \int_0^L dx \int_0^{\delta_1}  dz \ u \, w \, \right| = 
\left| \int_0^L dx \int_0^{\delta_1} dz \int_0^z dz' \, \p{u}{z'}(x,z') \int_0^{z'} dz'' \, \p{w'}{z''} (x,z'') \right| 
\nonumber \\
& \quad \leq \int_0^L dx \int_0^{\delta_1} dz \int_0^z dz' \left|\p{u}{z'}(x,z')\right| \sqrt{z'} 
\sqrt{\int_0^{\delta_1}  dz'' \left( \p{w}{z}(x,z'') \right)^2 } \nonumber \\
& \quad \quad \leq \int_0^L dx  \int_0^{\delta_1} dz \, \frac{z}{\sqrt{2}} \ \sqrt{\int_0^{\delta_1}  dz' \left( \p{u}{z}(x,z') \right)^2 } \ \sqrt{\int_0^{\delta_1}  dz'' \left( \p{w}{z}(x,z'') \right)^2 } \nonumber \\
& \quad \quad \quad \leq \frac{\delta_1^2}{4\sqrt{2}}\left(\frac{1}{C}\left\|\p{u}{z}\right\|_{x(0,\delta_1)}^2
+ \ \frac{C}{2}\left\|\p{u}{x}\right\|_{x(0,\delta_1)}^2+ \ \frac{C}{2}\left\|\p{w}{z}\right\|_{x(0,\delta_1)}^2\right)
\label{calc}
\end{align}
for any $C > 0$.
Choosing $C=\sqrt{2}$ we deduce that
\begin{equation}
\left|\int_0^L dx \int_0^{\delta_1} dz \, u \, w \, \right| \leq \frac{\delta_1^2}{8} \|\nabla {\bf u}\|_{x(0,\delta_1)}^2.
\end{equation}

Then a precisely analogous analysis may be performed in the top boundary layer because although $u$ does not (necessarily) vanish when $z=1$, $w$ does so the product $uw$ does.
Indeed, the computation in (\ref{calc}) does {\it not} use $u|_{z=0}=0$ or any boundary condition on $u$ at all.
This is where the Neumann boundary conditions require a change in the analysis from Dirichlet conditions \cite{CDPROLA,CD1994}; in the latter case a properly scaling bound appears without invoking incompressibility, but in the former case incompressibility (appears) to be absolutely necessary. This difference reflects the fact that Neumann conditions do not permit us to bound $u$ with the norm of its derivative near the boundary.
This is a similar situation to that encountered in the fixed-flux vs. fixed temperature thermal convection case \cite{FixedFlux2002}.

Finally, setting $\delta_1=\delta_2=\delta$ we conclude that
\begin{equation}
\calq_U \geq \frac{1}{2} \|\nabla {\bf u}\|^2 - Gr \frac{\delta^2}{8} \left( \|\nabla {\bf u}\|_{x(0,\delta)}^2 + \|\nabla {\bf u}\|_{x(1-\delta,1)}^2 \right) > \left( \frac{1}{2} - Gr \frac{\delta^2}{8} \right) \|\nabla {\bf u}\|^2.
\end{equation}

Using this bound:
\begin{equation}
\frac{1}{2}\de{}{t}\|{\bf u}\|^2 \leq -\left(1-\frac{Gr\delta^2}{4}\right)\|\nabla {\bf u}\|^2+\left(\sqrt{L}Gr+\sqrt{2\delta}Gr\right)\|\nabla {\bf u}\|
\end{equation}

Since we may take $\delta$ as small as we would like, we set $\delta=0$,
\begin{align}
\frac{1}{2}\de{}{t}\|{\bf u}\|^2 &\leq -\|\nabla{\bf u}\|^2+\sqrt{L}Gr\|\nabla{\bf u}\| \\
&\leq -\left(\|\nabla{\bf u}\|-\frac{\sqrt{L}Gr}{2}\right)^2+\frac{L Gr^2}{4}
\end{align}
We invoke the Poincare inequality, which for functions satisfying Dirichlet boundary conditions $z=0$, and Neumann boundary conditions
at $z=$, is $\|\nabla {\bf u}\|\geq \frac{\pi}{2}\|{\bf u}\|$. If $\|{\bf u}\|>\frac{\sqrt{L}Gr}{\pi}$, then we can use the Poincare inequality inside
the squared term:
\begin{equation}
\frac{1}{2}\de{}{t}\|{\bf u}\|^2 \leq -\left(\frac{\pi}{2}\|{\bf u}\|-\frac{\sqrt{L}Gr}{2}\right)^2+\frac{L Gr^2}{4}
\end{equation}

Using this inequality, if $\|{\bf u}\|\geq \frac{2\sqrt{L}Gr}{\pi}$, then $\de{}{t}\|{\bf u}\|^2\leq 0$. Therefore $\|{\bf u}\|$ is uniformly bounded by $\frac{2\sqrt{L}Gr}{\pi}$, and
the time averaged expression (\ref{thing}) is justified.

Having established uniform boundedness of the kinetic energy, we switch gears and prove bounds on the friction coefficient.
Substitute ${\bf i} U(z)+\bar{\tilde{{\bf u}}}(z)=\bar{{\bf u}}(z)$ into $\left<|\nabla {\bf u}|^2\right>$ and take
a linear combination of the expansion with (\ref{thing}) eliminating the $\langle\p{\tilde{u}}{z}\de{U}{z}\rangle$ term to establish
\begin{align}
\frac{1}{2}\left<|\nabla {\bf u}|^2\right> &=-\left<\frac{1}{2}|\nabla\tilde{{\bf u}}|^2+\de{U}{z}
\tilde{u}\tilde{w}\right>+Gr(\bar{u}(1)-U(1))+\frac{1}{2}
\left<\left(\de{U}{z}\right)^2\right>
\end{align}
Now define the quadratic form 
\begin{equation}
Q_U\{\tilde{{\bf u}}\} = \left<\frac{1}{2}|\nabla\tilde{{\bf u}}|^2+\de{U}{z}\tilde{u}\tilde{w}\right>, 
\end{equation}
which is the time average of $\calq_U$,
and use $\left<|\nabla {\bf u}|^2\right>=Gr \, \bar{u}(1)$ to deduce
\begin{align}
\bar{u}(1) &= 2U(1)-\frac{1}{Gr}\left<\left(\de{U}{z}\right)^2\right>+\frac{2}{Gr} \, Q_U.
\end{align}

Here is the essence of the background method: if we can choose $U(z)$ so that $Q_U$ is a non-negative quadratic form, then we have a lower bound for $\bar{u}(1)$ of the form
\begin{align}
\bar{u}(1) &\geq 2U(1)-\frac{1}{Gr}\left<\left(\de{U}{z}\right)^2\right>.
\end{align}
The task is to produce a background profile $U(z)$---subject to its boundary conditions---with $Q_U \ge 0$ producing as large a value of $2U(1)-\frac{1}{Gr}\left<\left( dU/dz \right)^2\right>$ as possible.

This may be accomplished by considering the same piece-wise linear background velocity profiles as used in the above demonstration of
uniform boundedness of the the kinetic energy.
Then $U(1)=Gr \, (\delta_1+\delta_2)$ and $\bar{u}(1)\geq Gr \, (\delta_1+\delta_2)$ when $Q_U \ge 0$. 
Hence the goal is to choose $\delta_1$ and $\delta_2$ to maximize their sum while keeping $Q$ non-negative definite.
Using the calculations that led to \ref{Qd22} leads to the equivalent expression:

\begin{equation}
Q_U \geq \left( \frac{1}{2} - Gr \frac{\delta^2}{8} \right) \|\nabla {\bf u}\|^2.
\end{equation}
This is positive if $\delta \leq 2 Gr^{-1/2} \leq \frac{1}{2}$ so $\bar{u}(1) \geq 4 Gr^{1/2}$ when $Gr \geq 16$.
In dimensional quantities this means $\bar{u}(h) \geq 4 \tau^{1/2}$, and the friction coefficient $C_f=\frac{\tau}{\bar{u}(h)^2}\leq\frac{1}{16}=.0625$ when $Re = \textrm{(non dimensional)} \ \bar{u}(1) \geq 16$.

\section{Higher $Gr$ bounds in three dimensions}

Much of the same algebra may be used to derive bounds for the three dimensional case: the strategy is the same except that there is a $y$ component in the velocity field that influences details of the estimates. 
The same bound, $\bar{u}(h)\geq 2U(1)-\frac{1}{Gr}\left< ( dU/dz )^2 \right>$, holds as long as
\begin{align}
Q_U = \frac{1}{2}\|\nabla {\bf u}\|^2 &+ Gr \int dx \int dy \int_0^{\delta_1}  dz \ u(x,y,z) \, w(x,y,z) \nonumber \\
&+ Gr  \int dx \int dy \int_{1-\delta_2}^1  dz \ u(x,y,z) \, w(x,y,z) \ > \ 0. \label{Qd3}
\end{align}
We restrict attention to the same two-parameter ($\delta_1$, $\delta_2$) background profile as in (\ref{background}) and Figure \ref{fig:background}.
In this case the boundary layers thicknesses will not be chosen to be equal.

We make the definition:
\begin{equation}
\|h\|^2_{xy(a,b)}=\int dx \int dy \int_a^b dz \,  h(x,y,z)^2, \label{note1}
\end{equation}
a generalization of the notation introduced above in (\ref{note}).
Beginning with the second term in $Q_U$ on the right hand side of (\ref{Qd3}) we use the fact that both $w|_{z=0}=0$ and $u|_{z=0}=0$ to deduce
\begin{center}
$$\left|\int dx \int dy \int_0^{\delta_1}dz\ u(x,y,z) \ w(x,y,z) \right|  \le$$ \\
$$\leq\int dx \int dy \int_0^{\delta_1}dz \, z \, \sqrt{\int_0^{\delta_1}dz'\left(\p{u}{z}(x,y,z')\right)^2  \int_0^{\delta_1}dz''\left(\p{w}{z}(x,y,z'')\right)^2 }$$ \\
$$\leq  \ \frac{\delta_1^2}{4}\left(\frac{1}{C} \left\|\p{u}{z}
\right\|_{xy(0,\delta_1)}^2 + \ \frac{C}{2} \ \left\|\p{w}{z}\right\|_{xy(0,\delta_1)}^2 +   \ 
\frac{C}{2} \ \left\|\p{u}{x}+\p{v}{y}\right\|_{xy(0,\delta_1)}^2\right) $$ \\
$$ \quad  \le \ \frac{\delta_1^2}{4}\left(\frac{1}{C}\left\|\p{u}{z}\right\|_{xy(0,\delta_1)}^2+
\frac{C}{2} \left\|\p{w}{z}\right\|_{xy(0,\delta_1)}^2 + \frac{C}{2} \left\|\p{u}{x}\right\|_{xy(0,\delta_1)}^2 \right. $$ \\
$$ \quad  \quad  \quad \left. + \ \frac{C}{2} \left\|\p{v}{y}\right\|_{xy(0,\delta_1)}^2 + \frac{C}{2} \left\|\p{u}{y}\right\|_{xy(0,\delta_1)}^2 + \frac{C}{2} \left\| \p{v}{x} \right\|_{xy(0,\delta_1)}^2 \right)$$
\end{center}

Choosing $C=\sqrt{2}$ we conclude
\begin{equation}
\left|Gr\int dx \int dy \int_0^{\delta_1} dz \, u\, w \right|
\leq\frac{Gr\delta_1^2} {4\sqrt{2}}\|\nabla {\bf u}\|_{xy(0,\delta_1)}^2.
\end{equation}

Bounding the term from the upper boundary layer in three dimensions is slightly more involved than in two.
First note that incompressibility implies
\begin{align}
& \int dx \int dy \int_{1-\delta_2}^{1} dz \ u(x,y,z) \ w(x,y,z)  = \nonumber \\
&\quad \quad = \ - \int dx \int dy \int_{1-\delta_2}^{1} dz \int_{z}^{1} dz' \ \p{u}{z}(x,y,z') \ w(x,y,z') \ \ + \nonumber \\
&\quad \quad \quad \quad  + \int dx \int dy \int_{1-\delta_2}^{1} dz \int_{z}^{1} dz'  \ u(x,y,z') \ \p{v}{y}(x,y,z'). \label{48}
\end{align}

The first term on the right hand side in (\ref{48}) above is estimated using only the fact that $w|_{z=1}=0$:
\begin{align}
&  \quad \quad \left| \int dx \int dy \int_{1-\delta_2}^{1} dz \int_{z}^{1} dz' \ \p{u}{z}(x,y,z') \ w(x,y,z') \right| \nonumber \\
& \le \int dx \int dy \int_{1-\delta_2}^{1} dz \int_{z}^{1} dz' \ \left|\p{u}{z}(x,y,z')\right| \sqrt{1-z'} \ \sqrt{ \int_{1-\delta_2}^{1} dz'' \left(\p{w}{z}(x,y,z'')\right)^2} \nonumber \\
&\le \int dx \int dy \int_{1-\delta_2}^1 dz \, \frac{1-z}{\sqrt{2}} \, \sqrt{ \int_{1-\delta_2}^1 dz' \left(\p{u}{z}(x,y,z')\right)^2
\int_{1-\delta_2}^1 dz'' \left(\p{w}{z}(x,y,z'')\right)^2 } \nonumber \\
& \quad \quad \quad \quad \le \frac{\delta_2^2}{2 \sqrt{2}} \ \left\|\p{u}{z}\right\|_{xy(1-\delta_2,1)} \left\|\p{w}{z}\right\|_{xy(1-\delta_2,1)}^2. \label{first}
\end{align}

The second term on the right hand side of (\ref{48}) requires a different approach.
Using only the fact that $u$ vanishes at the (relatively distant) bottom boundary, the inner integral may be bounded according to
\begin{align}
&\left| \int_{z}^{1} dz'  \ u(x,y,z') \ \p{v}{y}(x,y,z') \right|^2 \le
\int_{1-\delta_2}^{1} dz'  \ u(x,y,z')^2  \int_{1-\delta_2}^{1} dz''  \left(\p{v}{y}(x,y,z'')\right)^2 \nonumber \\
& \quad \quad \quad \le  \int_{1-\delta_2}^{1} dz' \, z' \, \int_0^{z'}  dz'' \left(\p{u}{z}(x,y,z'')\right)^2 \int_{1-\delta_2}^{1} dz''' \left(\p{v}{y}(x,y,z''')\right)^2 \nonumber \\
&  \le \delta_2 \left(1 - \frac{\delta_2}{2} \right) \int_0^{1}  dz'' \left(\p{u}{z}(x,y,z'')\right)^2 \int_{1-\delta_2}^{1} dz''' \left(\p{v}{y}(x,y,z''')\right)^2
\end{align}
Thus
\begin{align}
&\left| \int dx \int dy \int_{1-\delta_2}^{1} dz \int_{z}^{1} dz'  \ u(x,y,z') \ \p{v}{y}(x,y,z') \right| \nonumber \\
& \quad \quad \le \delta_2^{3/2} \left\|\p{u}{z}\right\|_{(0,1)} \left\|\p{v}{y}\right\|_{xy(1-\delta_2,1)} . \label{second}
\end{align}

Combining (\ref{first}) an (\ref{second}) we conclude
\begin{align}
& \left| Gr  \int dx \int dy \int_{1-\delta_2}^{1} dz \ u(x,y,z) \ w(x,y,z) \right|  \le \nonumber \\
&\quad \quad \le \ Gr \frac{\delta_2^2}{4} \ \|\nabla {\bf u}\|^2_{xy(1-\delta_2,1)} \ +
\ Gr \delta_2^{3/2} \left\|\p{u}{z}\right\|_{(0,1)} \left\|\p{v}{y}\right\|_{xy(1-\delta_2,1)}
\end{align}
and that $Q_U > 0$ when
\begin{align}
Gr \left( \frac{\delta_1^2}{4\sqrt{2}} \|\nabla {\bf u}\|^2_{xy(0,\delta_1)} + \frac{\delta_2^2}{4} \|\nabla {\bf u}\|^2_{xy(1-\delta_2,1)} 
+ \frac{\delta_2^{3/2}}{2} \|\nabla {\bf u}\|^2_{(0,1)} \right) < \frac{1}{2} \|\nabla {\bf u}\|^2.
\end{align}
Hence we may choose $\delta_1 < 2^{3/4} Gr^{-1/2}$ and $\delta_2$ arbitrarily small to establish the bound $\bar{u}(1)\geq 2^{3/4}  Gr^{1/2}$ and $C_f\leq\frac{1}{2\sqrt{2}}=.35355\dots$, displaying the same scaling albeit with an order of magnitude larger prefactor than Tang {\it et al}'s numerical bound $C_f \lesssim \frac{(Re+20.31)^2}{56.71Re^2}$ \cite{TCY} . The latter bound is a variable bound that depends on $Re$, has a maximum value of $.0231$ and tends asymptotically to $.01763$ as $Re$ goes to infinity. 
The results are plotted in Figure \ref{fig:stanton}.
The friction coefficient is bounded from below by $1/Re$, which we also plot in order to mark the range of accessible $C_f$ for $Gr>51.7$.

\begin{figure}
\begin{center}
\includegraphics[width = 2.8in]{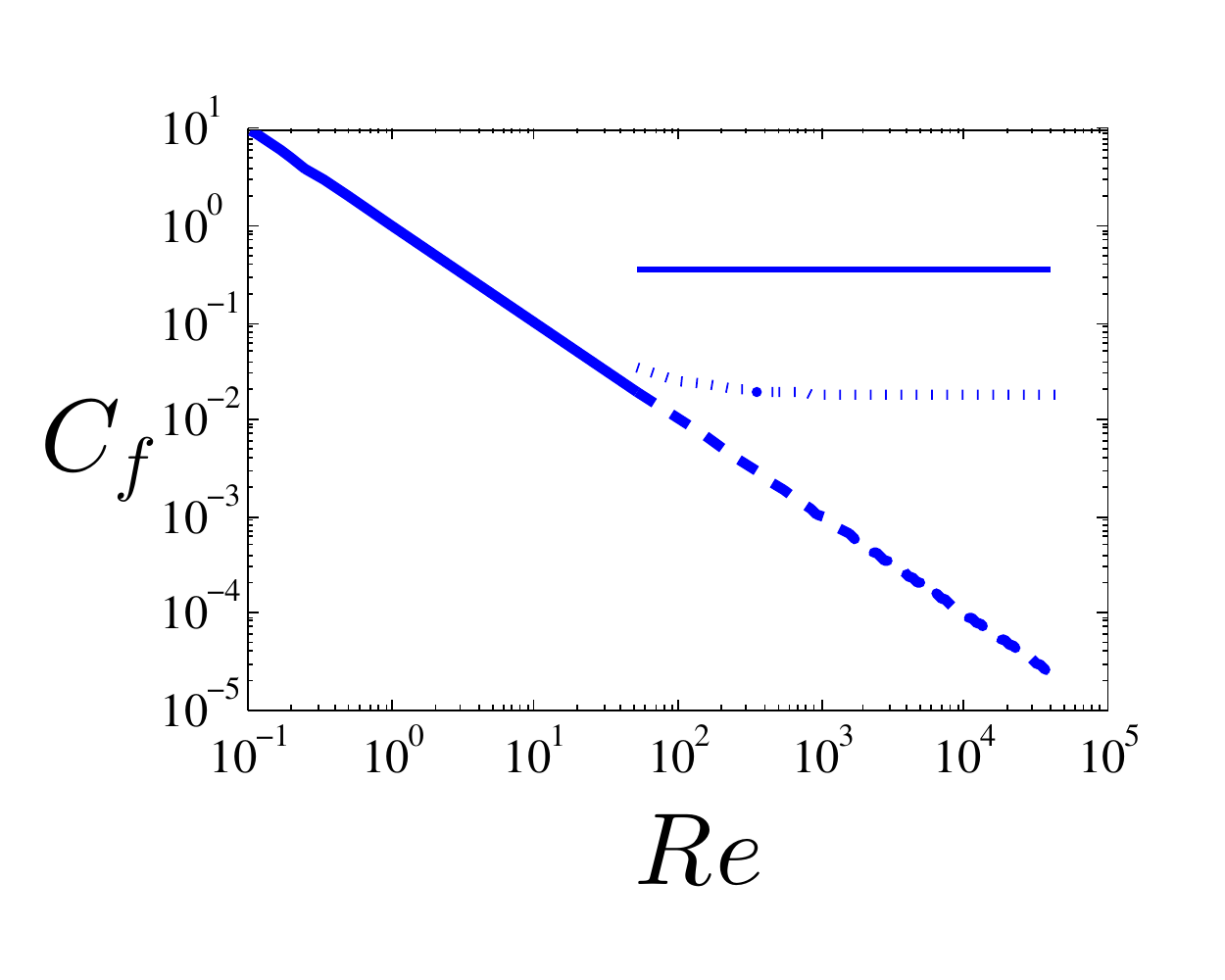}
\end{center}
\caption{Friction coefficient $C_f$ in terms of $Re$. The transition between solid and dashed diagonal lines indicates our lower bound on the transition between stable and unstable laminar flow.
The dotted horizontal curve $C_f=\frac{(Re+20.31)^2}{56.71Re^2}$ is the bound on the friction coefficient computed numerically \cite{TCY}, and the solid horizontal line $C_f=.35355\dots$ is the bound proved here.}
\end{figure}\label{fig:stanton}

\section{Acknowledgements} 

The authors gratefully acknowledge the hospitality of the Geophysical Fluid Dynamics Program at Woods Hole Oceanographic Institution, supported by NSF and ONR, where this work was begun.
This work was also supported by in part by USDOE Award DE-FG02-ER53223 (GIH) and NSF Awards PHY-0555324, PHY-0855335, and PHY-1205219 (CRD).


\end{document}